\documentclass[review]{elsarticle}

\usepackage{lineno,hyperref}
\usepackage{tipa}
\usepackage{amsmath}
\usepackage{amssymb}
\usepackage{setspace}
\usepackage{algorithm}
\usepackage{algpseudocode}
\usepackage{subcaption}
\usepackage{graphicx}
\usepackage{booktabs}
\usepackage{bm}
\usepackage{tabularx}

\usepackage{float}

\modulolinenumbers[5]

\journal{Speech Communication}

\biboptions{authoryear}

\begin{document}

\begin{frontmatter}

\title{The Mason-Alberta Phonetic Segmenter: A forced alignment system based on deep neural networks and interpolation}
\tnotetext[mytitlenote]{Fully documented templates are available in the elsarticle package on \href{http://www.ctan.org/tex-archive/macros/latex/contrib/elsarticle}{CTAN}.}


\author[gmuaddress]{Matthew C. Kelley\corref{mycorrespondingauthor}}
\ead{mkelle21@gmu.edu}

\author[uabaddress]{Scott James Perry}

\author[nauaddress,uabaddress]{Benjamin V. Tucker}

\address[gmuaddress]{Linguistics Program, Department of English, George Mason University, Fairfax, Virginia, 22030}

\address[uabaddress]{Department of Linguistics, University of Alberta, Edmonton, Alberta, T6G 2E7, Canada}

\address[nauaddress]{Department of Communication Sciences and Disorders, Northern Arizona University, Flagstaff, Arizona, 86011}


\cortext[mycorrespondingauthor]{Corresponding author}

\begin{abstract}
Forced alignment systems automatically determine boundaries between segments in speech data, given an orthographic transcription. These tools are commonplace in phonetics to facilitate the use of speech data that would be infeasible to manually transcribe and segment. In the present paper, we describe a new neural network-based forced alignment system, the Mason-Alberta Phonetic Segmenter (MAPS). The MAPS aligner serves as a testbed for two possible improvements we pursue for forced alignment systems. The first is treating the acoustic model in a forced aligner as a tagging task, rather than a classification task, motivated by the common understanding that segments in speech are not truly discrete and commonly overlap. The second is an interpolation technique to allow boundaries more precise than the common 10 ms limit in modern forced alignment systems. We compare configurations of our system to a state-of-the-art system, the Montreal Forced Aligner. The tagging approach did not generally yield improved results over the Montreal Forced Aligner. However, a system with the interpolation technique had a 27.92\% increase relative to the Montreal Forced Aligner in the amount of boundaries within 10 ms of the target on the test set. We also reflect on the task and training process for acoustic modeling in forced alignment, highlighting how the output targets for these models do not match phoneticians' conception of similarity between phones and that reconciliation of this tension may require rethinking the task and output targets or how speech itself should be segmented.
\end{abstract}


\end{frontmatter}

\section{Introduction}

Many speech-related research and engineering tasks require fine-grained time-alignment of segments (e.g., phones and phonemes) and words. For small amounts of data, this alignment task can be performed manually by domain experts like phoneticians. However, as we can attest, the time it takes to manually transcribe and align larger data sets becomes infeasible without using many annotators, who themselves must be well-trained in the practice of phonetic alignment. Indeed, \citet[][Chapter 9]{jurafsky_speech_2009} estimated that it can take 400 hours to phonetically label just 1 hour of speech, which is obviously prohibitive for large amounts of speech.

Automatic phonetic alignment methods are often used in lieu of manual methods to align segment and word boundaries to acoustic data. A phonetic alignment system must take in a recording and produce a time-aligned segmentation of the recording. The most common method of phonetic alignment is forced alignment, where an acoustic model is used to produce the segmentation based on phone or phoneme probabilities. The present study highlights some of the theoretical shortcomings of current approaches to developing forced alignment tools and proposes an alternative method using interpolation and category tagging that ameliorates some of the theoretical issues of previous approaches.

\begin{figure}[h]
	\begin{subfigure}{\textwidth}
 \includegraphics{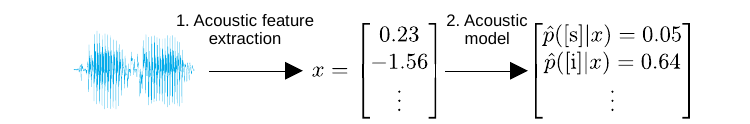}
	\subcaption{Conversion from acoustics to posterior probabilities.}
	\vspace*{0.25in}
	\end{subfigure}
 \begin{subfigure}{\textwidth}
 \includegraphics[width=\textwidth]{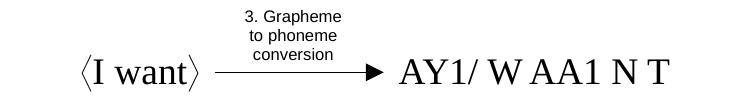}
	\subcaption{Conversion of graphemes to phonemes.}
	\vspace*{0.25in}
\end{subfigure}
\begin{subfigure}{\textwidth}
	 \includegraphics[width=\textwidth]{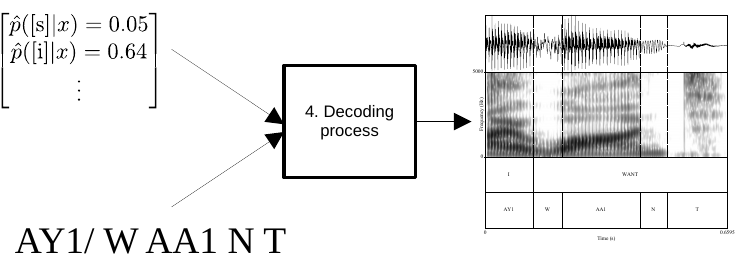}
	 \subcaption{Generation of alignment from probabilities and transcription.}
	 \end{subfigure}
 \caption{Flowchart diagram of forced alignment process. The (a) and (b) sections represent parallel streams that do not depend on each other. The output of (a) and (b) are then merged in (c) with the decoding process, which yields an alignment that can be displayed with a spectrogram and/or waveform.}
 \label{fig:alignment_diagram}
\end{figure}

A forced alignment system takes two objects as inputs: a recording and an orthographic transcription of what was said in the recording. There are several working parts in a forced alignment system that process those two inputs, and these parts are visualized in Figure~\ref{fig:alignment_diagram}. The first is a feature extractor that operates on the recording to extract useful lower-dimensional features from the highly multidimensional speech signal. The second is an acoustic model that assigns posterior probabilities over different segment categories (often phones or phonemes), given the features from the feature extractor. The third part is an orthography conversion system (like grapheme-to-phoneme models) that will convert orthographic transcriptions to phonemic transcriptions. These conversion systems often take the form of a look-up dictionary like the CMU Pronouncing Dictionary \citep{lenzo_cmu_2013}. And, the fourth part is a decoder that will process the posterior probabilities to align the phonetic transcription to the extracted speech features in an optimally probable fashion. What makes a forced aligner ``forced'' is that the system is coerced to produce an alignment that matches the given transcription, even if other alignments may be more acoustically probable.

Forced alignment has often been used when working with what is considered large amounts of speech data in experimental phonetics. \citet{kiefte_modeling_2017}, for example, used the Penn Phonetics Lab Forced Aligner \citep{yuan_speaker_2008} to perform phonemic alignment on 35 hours of speech to be able to ultimately extract vowel formant contours from the speech signal. \citet{pitt_buckeye_2005} similarly used the Entropic Aligner \citep{wightman_aligner_1997} as a first-pass alignment step on 300,000 words-worth of conversational speech. The accuracy of forced alignment systems is not always high, however. \citeauthor{kiefte_modeling_2017} also reported training the CMUSphinx 3 aligner \citep{seymore_1997_1998} on their data, but it did not provide usable results, in contrast with the Penn Phonetics Lab Forced Aligner. As such, it is of paramount importance to evaluate the accuracy of forced alignment systems both while developing them and when researchers, engineers, or practitioners are applying them.

The acoustic models can be improved by using deep neural networks instead of hidden Markov models. \citet{kelley_comparison_2018} have already shown preliminary results demonstrating improvements in alignment for the TIMIT speech corpus \citep{garofolo_darpa_1993}. Many of the current forced alignment systems that are used in modern research utilize hidden Markov models \citep{mcauliffe_montreal_2017,yuan_speaker_2008,gorman_prosodylab-aligner:_2011,kisler_signal_2012}. The relatively recent, marked improvement in acoustic modeling by deep neural networks instead of hidden Markov models \citep[as remarked on in, e.g.,][]{hinton_deep_2012} should lead to better performance in the alignment task \citep{kelley_comparison_2018}.

Another shortcoming in current forced aligner systems is the level of granularity they provide in their output. It is standard to advance frames by 10 ms when calculating acoustic features for speech recognition like MFCCs. However, this means that the boundaries can only be placed at 10 ms intervals, which may be too coarse for many speech-related purposes \citep{tucker_spontaneous_2023}. \citet{kelley_comparison_2018} used a 1 ms frame advance instead for greater precision, but using smaller frame advances increases the size of the training data, thereby significantly increasing both the time it takes to train the model and the time it takes to make predictions from the model. We believe that an alternative to using shorter frame advances is to interpolate between the discrete frames during the alignment process, which would lead to more granular boundaries in the final alignment.

There are more shortcomings that we feel it appropriate to discuss here for fullness, but we will not be exploring solutions to them in the present paper. The first of these is that relying on a dictionary for grapheme-to-phoneme conversion limits how automatic the forced alignment can be. Any word (or similar) in the transcription not present in the dictionary will have to be added to the dictionary before the alignment process can take place. Second, dictionary usage limits the type of alignment that can be performed to the mostly phonemic level. While both of these points can be alleviated somewhat by using more allophonic detail \citep{dicanioUsingAutomaticAlignment2013} or a statistical grapheme to phoneme (or phone) model, there persists a risk that the output phonemic/phonetic form will not match what is actually said in the recording. This mismatch is especially true for more conversational registers of speech, where phonetic reduction may be more extreme than in more careful speech \citep{warnerPhoneticVariabilityStops2011}. The mismatch may also lead to a fourth problem of catastrophic misalisgnment, where a phoneme is present in the transcription but has no acoustic presence in the recording (or vice-versa), causing a cascade of misaligned phoneme labels that must be hand-corrected \citep[as discussed in][]{kiefte_modeling_2017}. Spontaneous catastrophes of misalignment do also occur without mismatches between transcription and utterance, but this is a general problem with forced alignment. A fifth limitation is that forced aligners have been found to produce more errorful results when dealing with different speaking rates \citep{baileyAutomaticDetectionSociolinguistic2016}. 

\subsection{Related work}

There are several existing forced alignment tools that have seen use in the speech research community. The Penn Phonetics Lab Forced Aligner \citep{yuan_speaker_2008} has seen continued use since it was made available. It was trained on the SCOTUS corpus and relies on a traditional hidden Markov model and Gaussian mixture model (HMM-GMM) combination. It does not provide an interface to be retrained on custom data. The ProsodyLab Aligner \citep{gorman_prosodylab-aligner:_2011} is similar, though it does provide a training interface that facilitates aligning non-English data or customization of models and was trained on a mix of laboratory speech and speech collected from across the internet. The MAUS Aligner \citep{wesenickApplyingSpeechVerification1994} and WebMAUS Aligner \citep{kisler_signal_2012} use an HMM-GMM approach and support a variety of languages. The aligner provided in LaBB-CAT \citep{fromontLaBBCATAnnotationStore2012} also uses HMM-GMM models. \citet{pengPracticalWayImprove2021} also used an HMM-GMM model to demonstrate some impressive improvements to alignment on the TIMIT \citep{garofolo_darpa_1993} speech corpus.

The Montreal Forced Aligner \citep[MFA,][]{mcauliffe_montreal_2017} also uses a hidden Markov model and Gaussian mixture setup by default, using Kaldi \citep{povey_kaldi_2011} as the backend. However, it also allows for a neural network model to be used, specified through a Kaldi recipe. Like the ProsodyLab Aligner, the Montreal Forced Aligner provides a training interface to create new models based on new data. In addition, the Montreal Forced Aligner is arguably the most popular modern forced alignment system for speech research.

The Gentle aligner \citep{ochshorn_gentle_2017} uses a neural network backend from a Kaldi recipe. We note that the network architecture for the Kaldi recipes are not straightforward to access for these systems, especially if the designer has tweaked them at all. Specifying the acoustic model for an aligner in a deep learning framework would make the network architecture more easily understood and permit more configuration and customization beyond what is available through Kaldi. Another neural network-based aligner is Prak for aligning Czech data \citep{hanzlPrakAutomaticPhonetic2023}. It uses a relatively simple fully-connected neural network structure in PyTorch \citep{paszkePyTorchImperativeStyle2019} and will search through pronunciation variants. Their choice of network structure was well-motivated for their goals, though for alignment systems that are intended for general use, using layers that more naturally handle sequences is desirable. \citet{hanzlicekUsingLSTMNeurala} used recurrent long short-term memory layers in their acoustic model and used an iterative technique to improve their boundaries to achieve good segmentation performance.

There are several other approaches to phonetic alignment that could also be used. The first is a warping-based alignment. The transcription is used to synthesize a recording with known phone boundaries. The synthetic speech is then aligned to the input recording using dynamic time warping, and the produced nonlinear warping path is processed to determine the boundaries in the input recording. Praat \citep{boersmaPraatDoingPhonetics2023} and aeneas \citep{pettarinAeneas2020} provide functions that perform this type of phonetic alignment. This type of phonetic alignment obviates the need for an acoustic model, though the quality of the alignment will depend on how well the synthesized speech matches the original recording.

Another method that can be used for phonetic alignment is merely applying standard speech recognition techniques, such as those discussed in \citet[Chapters 9-10]{jurafsky_speech_2009}. Before end-to-end neural models that mapped acoustics directly onto letters or graphs, acoustic models like those used in forced alignment were used. (Forced alignment was, in fact, a crucial step in the embedded training routine used to develop speech recognition systems.) These models were decoded similarly to how forced aligners are, except that the Viterbi or $n$-best decoding would yield an optimal alignment to the most probable sequence of phones that corresponded to words in the system's lexicon. Alignment through speech recognition would save a lot of researcher time due to not needing orthographic transcriptions, and is provided through newer versions of WEBMAUS \citep{kislerSignalProcessingWeb2012} and the Montreal Forced Aligner. However, these types of alignment systems are vulnerable to word recognition errors that could mislabel the segments and words in the recording.

A final alternative method is using a semi-supervised learning techniques. The Charsiu aligner \citet{zhuPhonetoaudioAlignmentText2022} provides such an option, where an alignment can be created using a transformer network, without the need for providing an orthographic transcription. As with using full speech recognition models for alignment, this type of method has the potential of using ``incorrect'' segment labels. It is worth noting that the Charsiu aligner provides a more traditional framewise forced alignment method as well. 

\subsection{The present paper}

In the present paper, we introduce a new forced alignment system, the Mason-Alberta Phonetic Segmenter (MAPS). We use this system as a testbed for solutions we propose to increase the accuracy and precision of forced alignment. These proposals are to change the nature of the ``correct'' answer for the acoustic model to increase the accuracy and to use an interpolation technique to move boundaries to between acoustic frames. Since the Montreal Forced Aligner is trainable and has recently been found to outperform several different forced alignment systems \citep{gonzalezComparingPerformanceForced2020,mcauliffe_montreal_2017}, we compare the results of our system to it as well. The trainability aspect is important to ensure that the comparison we make is as even as possible because training on different data sets could produce substantially different alignment qualities.

One of the overall goals of the present paper is to reflect on how we in linguistics have conceptualized the structure of speech and the subsequent learning task we ask of neural networks. These reflections are not necessarily intended to be critical of the field's current understanding of how speech works. Rather, our intention is more of an instrumental one, to examine and critique the compatibility of how the field treats and understands speech with how we are training acoustic models for speech recognition, especially forced alignment. It is possible that these reflections could also be relevant to linguistic theory \textit{per se}, but we are not making that an explicit goal. We present both a theoretical and empirical analysis.

\section{Theoretical analysis}

While any of the four components involved in a forced aligner could affect the overall alignment quality, we focus on two specifically: the acoustic model and the decoding/alignment algorithm. The acoustic model provides some quantification of how well each segment category matches the acoustic input, and the alignment algorithm is how the neural network output is decoded to estimate where to place the boundaries between segments. The acoustic model typically takes the form of a segment classifier, and the alignment algorithm typically takes the form of the Viterbi algorithm or a simplified version of it.

\subsection{Segment classification}

Segment classification is a classic problem in automatic speech recognition, and it is the heart of the acoustic model used in forced alignment. Classifying acoustic data into segments in some fashion is also integral to many models of speech perception and spoken word recognition \citep[][\textit{inter alia}]{goldinger_echoes_1998, youTISKEasytousePython2018,luceRecognizingSpokenWords1998,klattSpeechPerceptionModel1979,norrisShortlistBayesianModel2008,mcclellandTRACEModelSpeech1986}, whether explicitly modeled or assumed to happen before the processes being modeled.

A probabilistic segment classifier's job is to predict the most probable segment given the acoustic data it is processing. Formally, such a classifier will return the result of

\begin{equation}
    \hat{\psi} = \operatornamewithlimits{arg \, max}_{\kappa\in k} P(\psi = \kappa \vert x)
\end{equation}

\noindent where $\hat{\psi}$ is the predicted segment classification, $\psi$ is the true identity of the segment, $k$ is the set segments in the language, $\kappa$ is an element in $k$, and $x$ is some kind of acoustic data (often a vector). For this type of classifier, it is also the case that $\sum_{\kappa \in k} P(\psi = \kappa \vert x) = 1$. The set of probabilities over all of the segments---that is, all $\kappa \in k$---is the posterior probability distribution of the classifier. In other words, the predicted segment is the segment from the language that has the highest posterior probability given the acoustic features.

\subsubsection{An ideal segment classifier}

It is helpful to imagine what an ideal classifier might look like. The most obvious behavior is that it selects the correct segment for vector $x$; that is, segments are accurately classified. We will refer to this as the \textsc{accurate classification criterion}. There is more specific behavior to be described in this ideal classifier, however, that would reflect current knowledge on segments. The behavior, for example, does not need to be an all-or-nothing assignment of probability since there are other behaviors that will still satisfy the \textsc{accurate classification criterion}.

We believe that an ideal segment classifier using acoustic features or cues would spread the probability over several different segments because they share common features \citep{internationalphoneticassociationHandbookInternationalPhonetic1999,roachReport1989Kiel1989,ladefogedReflectionsIPA1990,jakobson_preliminaries_1952}, such as acoustic cues \citep{stevensAcousticPhonetics1998,fantSoundFeaturesPerception1967}. For example, when $\psi = \text{[k]}$, the system should not only assign probability mass to [k] but also to [g] because both [k] and [g] are velar stops and share acoustic similarity. It would still need to be the case that [k] is the most probable segment category to satisfy the \textsc{accurate classification criterion}, but there are infinitely many probability distributions that would result in that behavior. More concretely, when $\psi = \text{[k]}$, it would be more ideal to have a distribution that assigns $P(\psi=\text{k} \vert x) = 0.65,\, P(\psi=\text{g}\vert x)=0.35$ instead of $P(\psi=\text{k} \vert x) = 1,\, P(\psi=\text{g}\vert x)= 0$. We will refer to this desideratum as the \textsc{similarity reflection criterion}, that the probability distribution should reflect the similarity (acoustic, featural, etc.) between the segments of a language. We believe it is reasonable to posit the \textsc{similarity reflection criterion} because the field refers to such concepts as voicing, place, and manner cues, which intrinsically express acoustic similarity between consonants.

The type of probability distribution that is produced when following the \textsc{similarity reflection criterion} inherently has high entropy, owing to how probability mass is spread between many different segment categories. Returning to the example of when $\psi = \text{[k]}$, the expository ideal distribution we gave would have a base-2 entropy of $H = -1 (0.65\log_2(0.65) + 0.35\log_2(0.35)) = 0.93$. Whereas, the all-or-nothing distribution would have a base-2 entropy of 0. The maximum base-2 entropy for a binary classification is 1, achieved when the posterior is a uniform distribution. Despite the high entropy in the example ideal distribution, this type of acoustic model would still satisfy the \textsc{accurate classification criterion}. In more complicated systems with full sets of segments within a language, the shared features between several segment types would still produce high entropy posterior distributions, spreading the probability mass over several different segment categories.

The entropy of this kind of distribution can also be observed from confusion data in segment perception, such as from identification tasks \citep{miller_analysis_1955} or gating tasks \citep{smitsUnfoldingPhoneticInformation2003, warner_tracking_2014}. \citet{norrisShortlistBayesianModel2008} made a similar observation about recognition probabilities about how all-or-nothing probabilities for segments are not realistic, given the phonetic ambiguity of the speech signal. Ultimately, it appears that satisfying the \textsc{similarity reflection criterion} would require a classifier to produce moderate- to high-entropy distributions.

In sum, we posit two criteria that would be satisfied by an ideal classifier: 1) the \textsc{accurate classification criterion} and 2) the \textsc{similarity reflection criterion}.

\subsubsection{Training classifiers in practice}

Current acoustic models are trained using one-hot vectors as all-or-nothing target probability distributions, where the ``true'' segment category is assigned a probability of 1, and all other categories are assigned a probability of 0. In addition, this type of distribution has minimal entropy equal to exactly 0. Notwithstanding results and theory in phonetics suggesting a lack of complete separability between traditional speech sound categories \citep[][\textit{inter multa alia}]{ladefogedInformationConveyedVowels1957,magnusonEARSHOTMinimalNeural2020,miller_analysis_1955,smitsUnfoldingPhoneticInformation2003,norrisShortlistBayesianModel2008}, it is conceivable that this type of network could satisfy the \textsc{accurate classification criterion} if it worked perfectly. However, using one-hot vectors as the training targets will impel the network to try to assign as much probability mass to a predicted segment as possible, rather than spread it between similar sounds, which does not satisfy the \textsc{similarity reflection criterion}.

The acoustic model is also unlikely to learn features that are familiar to phoneticians. Rather, we believe that the network would learn features that maximize the discriminability of the segment categories in the posterior distribution and, thus, minimize the entropy of the posterior distribution. This type of minimal entropy posterior would not satisfy the \textsc{similarity reflection criterion} very well, owing to the discrepancy in entropy.

\begin{figure}
    \begin{subfigure}{\linewidth}
        \centering
        \includegraphics[width=\linewidth]{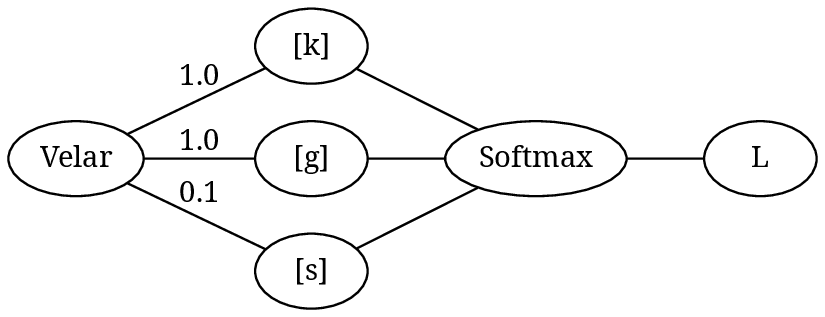}
        \subcaption{Initial state of network.}
        \label{fig:velarA}
    \end{subfigure}
    \begin{subfigure}{\linewidth}
        \centering
        \includegraphics[width=\linewidth]{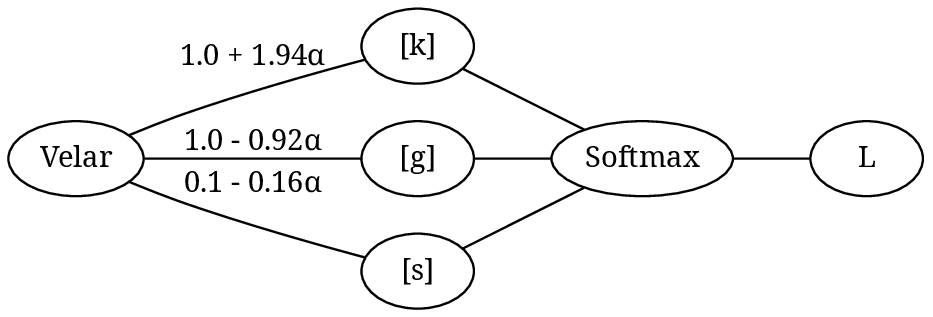}
        \subcaption{Network during backpropagated weight update.}
        \label{fig:velarB}
    \end{subfigure}
    \begin{subfigure}{\linewidth}
        \centering
        \includegraphics[width=\linewidth]{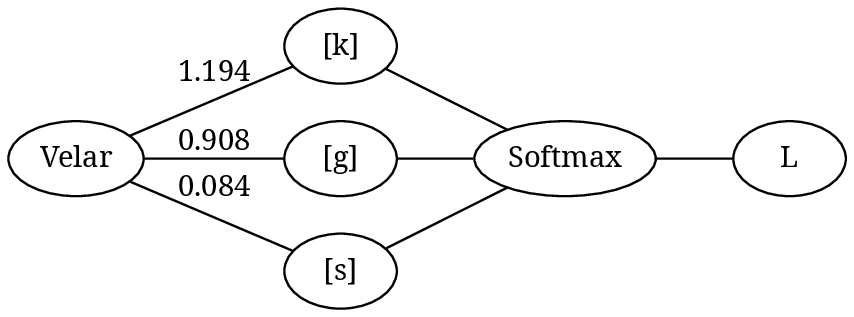}
        \subcaption{Network after gradient update.}
        \label{fig:velarC}
    \end{subfigure}
    \caption{Schematic state diagram of neural network with softmax activation and categorical cross-entropy loss when presented with exemplar labeled as [k]. The initial state of the network is shown in (a). In (b), a gradient update for vanilla stochastic gradient descent is shown for when the value of the ``Velar'' variable was 2. The numbers are derived using the gradient rules from Equations~\ref{eq:cce_corr}~and~\ref{eq:cce_inc}, in addition to basic partial derivatives of products. In (c), the state of the network after the update is shown, assuming $\alpha=0.1$ for the learning rate. Note how the connection strength between the velar pinch and [g] has been weakened, which is undesirable.}
    \label{fig:velar_whole}
\end{figure}

The impelled training behavior of the network can be observed by considering how an update from categorical cross-entropy loss is backpropagated through the network if the network were using familiar features. We describe this first formally, and then represent an example in Figure~\ref{fig:velar_whole}.

The function $\operatorname{cce}(z, y)$ expresses the cateogrical cross-entropy loss between a neural network's logit output vector $z$ and the target one-hot vector $y$. The gradient of $\operatorname{cce}(z, y)$ with respect to the logit output for the positive or ``correct'' segment category is

\begin{equation}
    \frac{e^{\zeta_p}}{\sum_j^n e^{\zeta_j}} - 1 \,,
    \label{eq:cce_corr}
\end{equation}

\noindent where $\zeta_p$ is the logit output for the correct segment, $\zeta_j$ the $j$-th logit output from the network, and $n$ is the total number of segment categories being considered. The derivation of this gradient from the categorical cross-entropy function is given in \ref{app:cce}.

The gradient output of $\operatorname{cce}(z, y)$ with respect to the logit output for the $i$-th negative or ``incorrect'' segment category is

\begin{equation}
    \frac{e^{\zeta_i}}{\sum_j^n e^{\zeta_j}} \,
    \label{eq:cce_inc}
\end{equation}

\noindent where $\zeta_i$ is the logit output for the $i$-th incorrect segment, $\zeta_j$ the $j$-th logit output from the network, and $n$ is the total number of segment categories being considered. The derivation of this gradient is also given in \ref{app:cce}.

The value of Equation~\ref{eq:cce_corr} will always be negative, and the value of Equation~\ref{eq:cce_inc} will always be positive. Thus, when performing a gradient descent update, the connections to the target output class will be strengthened, and the connections to the non-target output classes will be weakened.

Turning to a more concrete conceptual example, consider again the scenario of a system predicting between [k] and [g]. If there were a node in the network corresponding to velar cues in the acoustic input, the network would only be able to strengthen the connection between the velar cues, and one of [k] or [g] would be obliged to weaken the connection strength to the other segment.

This scenario is illustrated in Figure~\ref{fig:velar_whole}. The network is presented with an acoustic input vector, and its associated target label is [k]. During the gradient update, the categorical cross-entropy loss signal that is backpropagated through the network causes the connection strength between [k] and the velar feature to be increased. Concomitantly, the connection strength between [g] and the velar feature is decreased, which is undesirable. The strength between the velar pinch feature and [s] is also decreased, but this is desirable behavior.

It is unlikely that a network trained with categorical cross-entropy loss and one-hot vectors would learn this exact velar pinch feature, precisely because it is not sufficiently discriminative. However, this example does illustrate the challenge of using one-hot training vectors while also trying to satisfy the \textsc{similarity reflection criterion}. Ultimately, this training process forces all possible dissimilarities to be treated equally. That is, [k] is treated as acoustically dissimilar to [s] as it is to [g], which is clearly false on its face, but also not supported in acoustic measurements as in \citet{mielkePhoneticallyBasedMetric2012}.

This similarity problem extends to all acoustically similar or confusable segment categories. It therefore seems, at best, improbable that using one-hot vectors and categorical cross-entropy loss would satisfy the \textsc{similarity reflection criterion}. It is also unlikely that the \textsc{accurate classification criterion} could be satisfied because the set of segment categories that exist in a language are not entirely acoustically separable. In part, this inseparability is due to how speech sounds are defined based on similarities with other sounds. But additionally, the perceptions of some sounds depend on previous context \citep{ladefogedInformationConveyedVowels1957} or noise conditions \citep{miller_analysis_1955}, and the mapping from acoustics to segments is one-to-many or many-to-many \citep{magnusonEARSHOTMinimalNeural2020} and thus only approximable with neural networks.

It may seem that an alternative to segment labels could increase the performance of the system. However, as previously alluded to, the International Phonetic Alphabet (IPA) symbols that linguists use as category labels like [k], [g], and [s] are shorthand notations for different similarity-class features. Changing the nature of the labels from segments to other categories such as features would thus simply reify the problems we have already outlined regarding the \textsc{similarity reflection criterion}. Identifying gestures as in motor theory \citep{liberman_motor_1985} or direct realism \citep{fowler_event_1986} would also not resolve this issue because gestures themselves also share many similar features and would produce high entropy posteriors.

\subsubsection{A potential solution: Tagging}

We believe that it is possible to recast the acoustic model so as to avoid the use of one-hot vectors as training targets for classification. The recasting that we propose is to use multilabel classification instead of single-category classification. This method would decouple the positive and negative classes by replacing the softmax layer with a sigmoid layer that is applied to each element of the output and use the sum of a binary cross-entropy calculated on each output from the sigmoid layer as the loss function. In doing so, the acoustic model is set up as calculating multiple separate binary classifiers with shared weights until the final layer. This method is more or less equivalent to what \citet{he_joint_2018} call a ``joint binary neural network.'' In theory, this type of network would be capable of outputting the higher entropy posteriors that would satisfy the \textsc{similarity reflection criterion}. However, the output of this type of network would not be a single posterior distribution. Rather, it would be a set of posterior distributions.

Other researchers have also attempted to treat segment recognition as a multi-label classification problem, though not in application to forced alignment as far as we know. \citet{pushpa_multi-label_2017} converted the phone recognition problem into a binary relevance multi-label classification problem for Tamil. To do so, they decomposed phones into phonetic features and trained separate classifiers for each feature. \citet{vasquez-correaPhonetToolBased2019} did something similar for Spanish when creating the Phonet classifier for phonological features. However, \citet{he_joint_2018} critiqued the binary relevance format for multi-label classification, observing that such solutions ignore dependencies between labels and require training separate classifiers for every label, greatly increasing the training time and the overall model size. \citet{brouwer_feed-forward_2004} used another approach to convert one-to-many relations by using fuzzy sets as the output. The particular method they used was seemingly designed more so with regression applications in mind than classification applications. Nevertheless, the output from a joint binary neural network is not very different from a fuzzy set, where the on/off probability of each label in the output set could be interpreted as the amount that that label is a member of the output set.

When the acoustic model is cast as a multilabel classification problem, the binary cross-entropy loss is used on sigmoid activations. The backpropagated loss signal with respect to any segment category's logit output $\zeta_i$ is

\begin{equation}
    \begin{cases}
    \frac{-1}{e^{\zeta_i}+1} &\mbox{if } \psi_i = 1 \\
    \frac{e^{\zeta_i}}{e^{\zeta_i}+1} &\mbox{if } \psi_i = 0 \,,
    \end{cases}
\end{equation}

\noindent where $\psi_i$ is the target output for a segment category. The derivation of this gradient from the binary cross-entropy definition is given in \ref{app:bce}.

It can be seen that the update is always negative for the case where $\psi_i = 1$ because $e^{\zeta_i} + 1$ is always positive. Similarly, the gradient is always positive for the case where $\psi_i = 0$ because all the terms are positive, and none of the applied operations will change the signs when applied on positive numbers. As such, the gradient updates will be applied such that the relevant connections for labels with an outcome of 1 will be strengthened, and the relevant connections for labels with an outcome of 0 will be weakened. When only one label is allowed to have a value of 1 per training example, this results in the same problem encountered with softmax and categorical cross-entropy. However, when all the relevant labels are allowed to be 1, the appropriate connection strengthenings and weakenings will be applied.

The open question regarding a joint binary neural network casting of the acoustic model is how to determine the additional segment category labels. There may be a grounded way to determine what classes will significantly overlap with each other. However, research into how best to calculate acoustic distance between segments in perceptually relevant ways is nascent \citep{mielkePhoneticallyBasedMetric2012, kelleyUsingAcousticDistance2022, kelleyAcousticDistanceAcoustic2022, bartelds_new_2020}. We instead implement an empirical method to determine which classes are confusable by training a traditional phoneme classifier and using its errors to determine the labels for the multilabel classification. We then train a new network to minimize the binary cross-entropy based on those error-derived labels.

This method is intended to satisfy the \textsc{similarity reflection criterion} by using shared features to predict sounds and permitting the features to be associated with multiple segment categories. The way that this method satisfies the \textsc{accurate classification criterion} is less straightforward, though. It is still possible in theory for this kind of network to satisfy the \textsc{accurate classification criterion} since the intended segment category-as-tag could still receive the highest output probability.

There is a similarity between this method and the distillation methods from \citet{hintonDistillingKnowledgeNeural2015}. A common form of these methods is often called a ``teacher-student'' network, where some form of the probability distribution from a traditionally trained network (the teacher) is used as the target output for a newly trained network (the student). However, within this training method, it remains unclear how to correct or adjust the errorful distributions to output the correct category, which is an unfortunate weakness for a system to be used in forced alignment. Element theory in phonology \citep{backleyIntroductionElementTheory2011} also bears some conceptual similarity to the tagging approach here. However, the ``elements'' in our tagging proposal are the full set of English phonemes in the training set and not an actual elemental breakdown of each of the sounds.

\subsection{Boundary placement precision}

The output of the acoustic model for an entire utterance yields posterior class probabilities over time, sometimes referred to as a phonetic posteriorgram \citep[e.g.][]{hazenQuerybyexampleSpokenTerm2009,zhaoUsingPhoneticPosteriorgram2019}. To estimate where the boundaries are between segments, an alignment algorithm based on dynamic programming---such as the Viterbi algorithm---is used to produce an optimal alignment between a phonetic or phonological transcription and the acoustic vectors over time. Acoustic vectors being passed into the acoustic model are generally discretely sampled from the input speech signal. A typical choice for (log) mel filterbanks and mel frequency cepstral coefficients (MFCCs) is to use 25 ms windows sampled every 10 ms.

Once the alignment has been created, a system is generally limited to placing boundaries between individual acoustic vectors. Since the signal is only sampled every 10 ms to calculate the acoustic vectors, the boundary placement precision is effectively limited to 10 ms. This is a problematic limit for forced alignment when intended for phonetic analysis because, for example, reduced speech sounds and short-lag burst releases may be shorter than 10 ms.

To increase the precision of the boundaries, there are several relatively straightforward options. The first of these is to sample the signal more frequently. \citet{kelley_comparison_2018} sampled every 1 ms, though this grows the size of the training data by an order of magnitude and substantially increases the training time for the model. The boundaries can be included as output tokens, as in the hidden Markov model setup used in \citet{yuanUsingForcedAlignment2018}, though this may only increase placement accuracy and not precision of boundaries in the 0 to 10 ms interval. The types of errors the aligner makes can also be analyzed and corrected for \citep{stolckeHighlyAccuratePhonetic2014}, though it would be desirable if the forced alignment system did not need to be corrected in the first place.

An alternative technique is to interpolate between time steps in the aligned signal to find a more precise time to create the boundary. This type of interpolation is relatively common in some fields that use dynamic programming, even being textbook-level techniques in areas like algorithmic planning \citep{lavallePlanningAlgorithms2006}. The general idea is that dynamic programming often operates over discretely sampled representations of continuous data, and interpolation can be used to estimate values between the sampled points. Boundary precision should be increased by using this technique because boundaries are no longer restricted to discrete intervals in the signal.

\section{Empirical analysis}

To evaluate the tagging-based approach to the acoustic model, we trained two types of acoustic models. The first of these is a standard segment classifier that used one-hot vectors as targets and categorical cross-entropy loss. We refer to this model as the ``crisp'' model because it was trained to make crisp, all-or-nothing classifications. The second of these models is a multi-label tagging-based model, as motivated previously. We refer to this model as the ``sparse'' model because its output target is a sparse vector of mostly zeroes.

For both of these models, we tested a linear interpolation method as a way to increase the precision of the boundary placement. There were, thus, four systems we designed to test: 1) a classifier with interpolation, 2) a classifier without interpolation, 3) a tagger with interpolation, and 4) a tagger without interpolation. The models were trained using Keras interface \citep{chollet2015keras} in TensorFlow v2.9.1 \citep{abadiTensorFlowLargeScaleMachine2016}. Model evaluation took place using TensorFlow v2.12.0.

\subsection{Training data}

Two annotated speech corpora were used to train the models. The first corpus was TIMIT \citep{garofolo_darpa_1993}, which is a standard speech corpus used to train speech recognition models. All of the data in TIMIT is recordings of read sentences, and the speakers are from a variety of dialect regions. The second corpus was the Buckeye speech corpus, a collection of sociolinguistic interviews with 40 Caucasian speakers from the Columbus, Ohio region \citep{pitt_buckeye_2005,pittBuckeyeCorpusConversational2007}.

Speakers 27, 38, 39, and 40 were held out from Buckeye to be part of the test set. The held-out speakers ensured that there was one male speaker younger than 30, one female speaker younger than 30, one male speaker older than 40, and one female speaker older than 40. Speaker 4 from Buckeye was held out to be part of the validation set. Of the remaining training data, an additional 5\% was randomly held out as additional validation data. The standard TIMIT test set was used, including the sentences designed to elicit dialectal variation.

Because the Buckeye recordings were many minutes long each, they were automatically processed into shorter sections. The boundaries between each section were the non-speech noises indicated in the transcriptions, such as laughter, silence, or the interviewer speaking. This processing strategy yielded phrases that corresponded to the speech between pauses (of any sort) in the conversation.

For each TIMIT sentence and Buckeye phrase, a series of MFCCs, deltas, and delta-deltas were calculated using the \texttt{python\_speech\_features} package \citep[][v0.6]{lyonsPythonSpeechFeatures2020}. The default parameter settings were used, yielding thirteen coefficients for each frame, with a window length of 25 ms and a step of 10 ms. The zeroth coefficient was also replaced with the log of the frame energy. Each MFCC vector was assigned a label based on the accompanying transcription. When the frame straddled a segment boundary, the segment that constituted the majority of the frame was used, defaulting to the prior segment in the case of a tie.

We used a relatively standard TIMIT segment label folding \citep[][Chapter 7]{gravesSupervisedSequenceLabelling2012} with the following exceptions. First, instead of folding stop closures into a single silence category, they were folded into their appropriate stop category. So, for example, a TIMIT label of [pcl] (indicating the closure of [p]) was folded into the regular [p] category; this choice was to ensure that there was parity between Buckeye and TIMIT labels since Buckeye does not have stop closures separated from releases. Second, we did not move the glottal stop into a garbage category. The foldings we performed for Buckeye can be found in \ref{app:buckfold}. All frames associated with non-speech sounds were discarded. We note that the Buckeye folding did not contain the [zh]$\rightarrow$[sh] mapping ([\textipa{Z}]$\rightarrow$[\textipa{S}] in IPA) that the TIMIT folding did; this does not affect the evaluability of the models, though.

\subsection{Model architecture}

The models we trained shared the same architecture. The input layer had a size of 39, corresponding to the MFCCs, deltas, and delta-deltas for each time step. There were three hidden bidirectional long short-term memory (LSTM) layers with 128 units (per direction) each and a dropout value of 0.5. For the standard segment recognizer model, the output layer was fully-connected with softmax activation and had 61 units corresponding to each of the segment categories in the TIMIT database (even though some of the categories were folded into similar ones). The output layer for the sparse, tagging model was analogous, except that a sigmoid activation was applied to each unit instead of softmax. The output layer was applied to each time step in the input data, as were the LSTM layers.

This model architecture was arrived at based first on results form \citet{graves_framewise_2005} where usable accuracy in framewise prediction models can be achieved using bidirectional LSTM layers with 93 units (per direction). We stacked this type of layer several times to increase depth and model capacity. We did not explore greater depths or layer widths because the aligner needs to be able to be run on an average user's computer, whether via GPU or CPU processing. Greater model depth or width decreases the probability of the model being runnable on an average computer.

The input data was centered and scaled using the \texttt{LayerNormalization} layer in \texttt{Keras} before being given to the input layer. The activation after each hidden layer was also centered and scaled using the same layer type after each hidden layer.

\subsection{Model training routine}

A batch size of 64 was used for training. Padding was applied to ensure uniform sequence length within a batch, and masking was applied to the padding in each batch. The single class recognition model was trained for 50 epochs with the Adam optimizer set to the default values in \texttt{Keras}, and the model and its results were saved after each epoch. Categorical cross-entropy was used as the loss function. The best model was selected as the saved model from the epoch with the highest validation accuracy.

The multilabel model was trained based on the best crisp model in each run. The new targets were determined by gathering all of the model predictions on the training set from the crisp model and selecting all labels that were at least as probable as the original target label from the one-hot encoding as label targets. So, if the original target for a time step was [\textipa{I}], but [i] and [\textipa{E}] were calculated as at least as likely as [\textipa{I}] at that time step, the new targets for that timestep would be [i], [\textipa{I}], and [\textipa{E}]. Following, we calculated how many labels each time step had on average, which was approximately 2.

After determining the new targets, freshly initialized versions of the networks were trained. They were identical to the previous crisp models except that the final layer had a sigmoid activation function. The loss function used this time was the \texttt{weighted\_cross\_entropy\_with\_logits} function from TensorFlow, and the logits were calculated based on the sigmoid activation output of the model. This loss function calculates the binary cross-entropy for each value in the output and applies a user-specified weight to the loss associated with positive cases in the target values. In our case, because there were an average of 2 labels per time step, we used a weight of 30 so that the positive labels would affect the loss function and gradient updates approximately as much as the negative labels would. When this kind of weighting is not done, the amount of 0s in the output greatly outnumber the number of 1s, so the network struggles to output positive cases.

As before, the network was saved after each epoch. Its loss, sensitivity, specificity, and balanced accuracy on both training and validation sets were saved after each epoch. The best performing network was chosen as the one with the lowest balanced accuracy on the validation set. As defined in \citet{brodersenBalancedAccuracyIts2010}, balanced accuracy is the average of the sensitivity and specificity of the model. In the context of phone recognition, this measure indicates both how often the model correctly tags an audio frame with a relevant phone label (sensitivity), in addition to how often the model correctly does not tag a frame with an irrelevant phone label (specificity).

\subsection{Decoding the network output}

We use a simplified and corrected version of the decoding algorithm that \citet{kelley_comparison_2018} used. The algorithm is effectively a modified dynamic time warping algorithm between the desired phoneme labels and time steps, where the cost function is the absolute value of the log of the network's outputs, and warping a single time step across multiple labels is disallowed. By virtue of being a dynamic programming algorithm, it is structurally and conceptually similar to the Viterbi algorithm as well. The corrected algorithm is presented in Algorithm \ref{alg:decode} and uses 1-based indexing. The algorithm accepts a matrix $O$ and a sequence $s$ as input, where the matrix $O$ is the absolute value of the log-transformed softmax or sigmoid output from the network, and $s$ is the length-$n$ sequence to align in time, represented as integers referring to the appropriate phone category in the network's output. $O$ has the shape $k \times T$, where $k$ is the number of phone categories and $T$ is the total number of discrete time steps in the input data.

\begin{algorithm}[!ht]
	\caption{Decoding neural network output.}
	\label{alg:decode}
\begin{algorithmic}[1]
\Function{Decode}{O, s}
	\State Initialize an $n+1$ by $T+1$ matrix $M$ filled with $\infty$ values
	\State $M[1,1] \gets 0$
	\For{$i$ from 2 to $n+1$}
		\For{$t$ from 2 to $T+1$}
            \State $d \gets O[s[i],t]$
            \State $M[i,t] \gets d + \min(M[i-1,t-1], M[i,t-1])$
		\EndFor
	\EndFor
    \State Remove the first row and first column of $M$
	\State \Return The path obtained by performing backtracking on $M$
\EndFunction
\end{algorithmic}
\end{algorithm}

The specific correction compared to \citet{kelley_comparison_2018} is that the arg max function used and reverse traversal of the resulting array \citeauthor{kelley_comparison_2018} described is not guaranteed to produce the optimal path, merely a path where the frames indices are non-decreasing. We note that this correction does not affect the results \citeauthor{kelley_comparison_2018} presented because the implementation of the algorithm did not make use of the arg max approach, where the arg max approach had been intended to simplify the code structure of the algorithm when presented in pseudocode in the resultant paper.

Once a path is determined from calling \textsc{Decode}, the boundaries are simply placed when the symbol in the path transitions to the next symbol, taking into account the window length and window step when calculating the acoustic features. Since we used a 25 ms window with a 10 ms step, the boundary between two adjacent sections of a particular symbol is calculated as

\begin{equation}
    t = 0.025 + (i-1)0.01 = 0.015 + 0.01i\,,
    \label{eq:bound_to_time}
\end{equation}

\noindent where $t$ is the time of the boundary in ms, and $i$ is the index in the path sequence where the transition is happening, assuming the sequence is indexed from 1. For example, if the index was 1, the boundary would be drawn at $0.015 + 0.01(1) = 0.25$ ms. And, if the index was 10, the boundary would be drawn at $0.015 + 0.01(10) = 0.115$ ms.

We also implemented an option for linear interpolation when determining the boundary. The motivating idea is that the transition points when backtracking in $M$ from Algorithm~\ref{alg:decode} can be treated as two functions that likely cross over at some point between the two discretized time points. We chose to use linear functions to model that crossover since they are straightforward to use and find intersection points for.

Let $A$ be the 2x2 matrix that corresponds to the submatrix in $M$ where backtracking would cause a transition between symbols. We can treat $A$ as the y-values of 4 points that determine two line segments that intersect. To find the intersection between these two points, we can use a unit increase (such as when an array index is increased by 1) as the x-values in the points. $A$ is structured as follows:

\begin{equation}
    A = \left [ \begin{array}{c|c}
        \alpha_{1,1} & \alpha_{1,2} \\
        \alpha_{2,1} & \alpha_{2,2}
        \end{array} \right ] \,,
\end{equation}

\noindent and we will refer to the columns of $A$ as $a_1$ and $a_2$. The column-vector of slopes $d$ (as deltas between the columns in $A$) is then $d = a_2 - a_1$.

Finally, we can create a system of equations. In slope-intercept form, we can treat $\alpha_1$ and $\alpha_3$ as intercepts, giving a general form of $\psi = \delta_i \chi + \alpha_j$. By subtracting $\delta_i \chi$ from both sides, the following system of equations is achieved:

\begin{align}
    -\delta_1\chi + 1\psi &= \alpha_{1,1} \\
    -\delta_2\chi + 1\psi &= \alpha_{2,1}\,,
\end{align}

\noindent It is then possible to solve for $\chi$ and $\psi$ using linear algebra or simple algebraic manipulation. The value of $\chi$ corresponds to how much additional time needs to be added to the boundary, as a proportion of the window length from the acoustic processing. Since we have chosen 10 ms as the window length, the additional time in seconds added to the boundary would be calculated as $0.010\chi$. The value of $\psi$ is the interpolated decoding score between the points in $A$, the value of which is irrelevant for our purposes.

In some circumstances, $\chi$ may not be on the interval $[0, 1]$, such as if the last symbol were a very bad fit for the acoustic frames in comparison to the penultimate symbol. It is also possible for there to be no intersection if the line segments are parallel, though we believe this scenario would be rare and require a long series of equiprobable outputs from the acoustic model and/or some kind of numerical precision overlap brought on by using finite representations of numbers. In either of these cases, the simple resolution is to not add any sort of interpolated value to the duration and just keep the value calculated using Equation~\ref{eq:bound_to_time}.

\subsection{Montreal Forced Aligner}

We also trained the Montreal Forced Aligner (v1.0.1) to provide a point of comparison. We intentionally chose to train it instead of using a pre-built model to ensure an even comparison that wasn't affected by the type and amount of data being trained on. To ensure that the phone sets being trained on were identical, each TIMIT sentence or Buckeye phrase was listed as a ``word'' in the pronunciation dictionary for the Montreal Forced Aligner to look up, and the associated pronunciation was the target transcription for the entire utterance.

The acoustic model was trained using the Montreal Forced Aligner's built-in function \texttt{train\_and\_align} with default settings, passing the same training data as the above models. We saved the resulting acoustic model and used it to align the validation and test sets. 

\subsection{Results}

It is important to lay out how we will evaluate our systems. We plan for MAPS to be useful for phoneticians and other speech researchers. As such, we note that at least a portion of the evaluations performed on the system must be relevant to the kinds of research that uses the boundaries placed by forced aligners. It is well-known in speech research that some speech events can occur over the span of only a few milliseconds. As previously mentioned, one example is voice onset time, which can average less than 10 ms in some languages \citep{liskerCrosslanguageStudyVoicing1964}. For this reason, it is crucial for estimates of boundary placement error to be reported in the evaluation, such as in the form of the absolute error of the boundaries.

It is indeed common for boundary error evaluation to be reported in comparisons of forced aligners \citep{kelley_comparison_2018, mcauliffe_montreal_2017, pengPracticalWayImprove2021, gonzalezComparingPerformanceForced2020}. However, many studies that do report some degree of boundary error do so in terms of percentages of boundaries that fall below a particular error threshold, such as the number of boundaries with errors below 20 ms \citep{mcauliffe_montreal_2017, pengPracticalWayImprove2021}. These thresholds give a sketch of how the alignment system performs at various levels of desired precision. However, a much fuller picture of the errors a system produces can be obtained by plotting the empirical cumulative density function (CDF) of the absolute errors. Because the error threshold percentages are simply a coarsely sampled empirical estimate of the CDF, using the CDF is a natural generalization.

Some alignment and segmentation systems did not report on the boundary errors themselves, though. Of these, it has been common to binarily report an absolute boundary error of less than or equal to 20 ms as accurate and anything else as inaccurate \citep{stolckeHighlyAccuratePhonetic2014, hoangBlindPhoneSegmentation2015}. Others have reported performance in terms of precision, recall, and related metrics based on some kind of pre-determined threshold \citep{zhuPhonetoaudioAlignmentText2022, michelBlindPhonemeSegmentation2017, kamperUnsupervisedPhoneWord2021}. We acknowledge that not all of these systems were developed with forced alignment---as used in phonetic and speech research---in mind. However, we remark that systems that this type of evaluation is not informative of how such systems will perform when phonetically segmenting speech. While there is assuredly some degree of correlation between metrics like precision and recall and an aligner's performance, the relationship is difficult to reason about. Additionally, while a 20 ms threshold for accuracy might be reasonable for some specific, long speech sounds like vowels, it is all but useless for short events like stops with short voice onset times, flaps, and highly reduced sounds.

\subsubsection{Crisp model evaluation}

\begin{figure}[h!]
    \centering
    \begin{subfigure}[t]{0.49\columnwidth}
        \centering
        \includegraphics[width=\columnwidth]{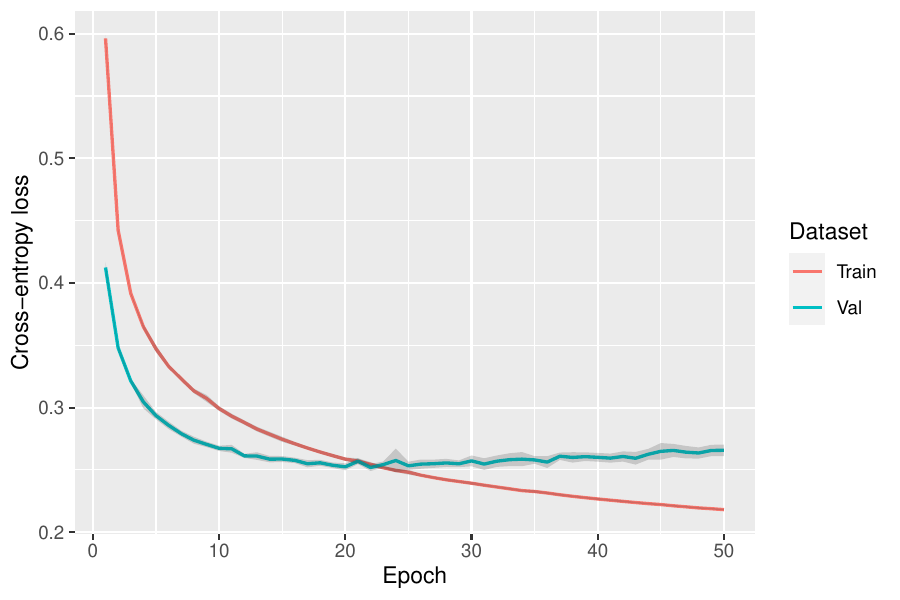}
        \caption{Cross-entropy loss for crisp model.}
        \label{fig:crisp_loss}
    \end{subfigure}
    \begin{subfigure}[t]{0.49\columnwidth}
        \centering
        \includegraphics[width=\columnwidth]{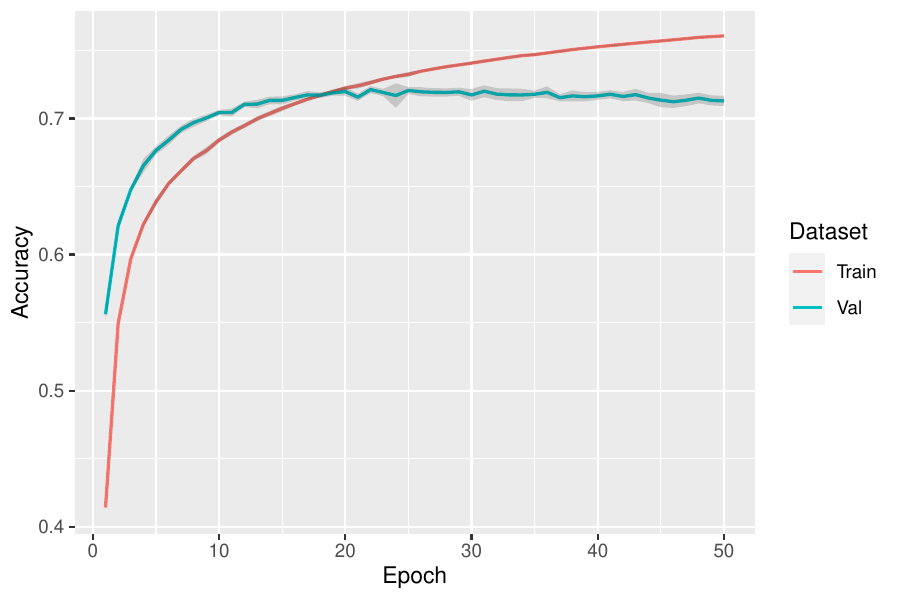}
        \caption{Accuracy for crisp model.}
        \label{fig:crisp_acc}
    \end{subfigure}
    \caption{By-epoch evaluation metrics for crisp network. Bands corresponding to a 95\% confidence interval are given in the shading surrounding the line plots. Note that the bands are present for the training metrics, but they are very tight.}
    \label{fig:crisp_metrics}
\end{figure}

During training, the crisp network was evaluated based on the accuracy of its predictions. The maximum validation accuracy was achieved after an average of 28.50 epochs ($SE = 2.43$). The overall best loss and accuracy are presented in Table~\ref{tab:crisp_metrics}. The by-epoch categorical cross-entropy loss and accuracy on the training set and the validation set can be seen in Figure~\ref{fig:crisp_metrics}. 

\begin{table}[h!]
    \centering
    \begin{tabular}{lcc}
        \toprule
         Data set & Loss & Accuracy  \\
         \midrule
         Train & $0.24 \pm 0.008$ & $0.74 \pm 0.008$ \\
         Validation & $0.25 \pm 0.002$ & $0.73 \pm 0.002$ \\
         Test & $1.01 \pm 0.009$ & $0.71 \pm 0.002$ \\
         \bottomrule
    \end{tabular}
    \caption{Evaluation metrics for the crisp network trained with one-hot target vectors. Each value is given as its mean across the best epochs of the ten models trained $\pm 1.96\,SE$ to indicate a 95\% confidence interval estimate. The loss in this case was standard categorical cross-entropy.}
    
    \label{tab:crisp_metrics}
\end{table}

\subsubsection{Tagging network evaluation}

\begin{figure}[h!]
    \centering
    \begin{subfigure}[t]{0.49\columnwidth}
        \centering
        \includegraphics[width=\columnwidth]{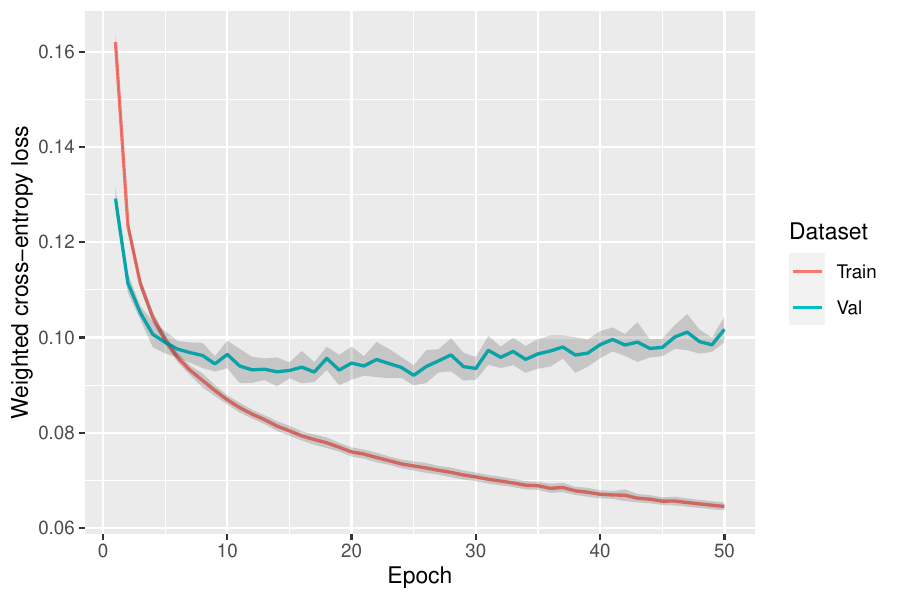}
        \caption{Weighted binary cross-entropy loss for tagger model.}
        \label{fig:tag_loss}
    \end{subfigure}
    \begin{subfigure}[t]{0.49\columnwidth}
        \centering
        \includegraphics[width=\columnwidth]{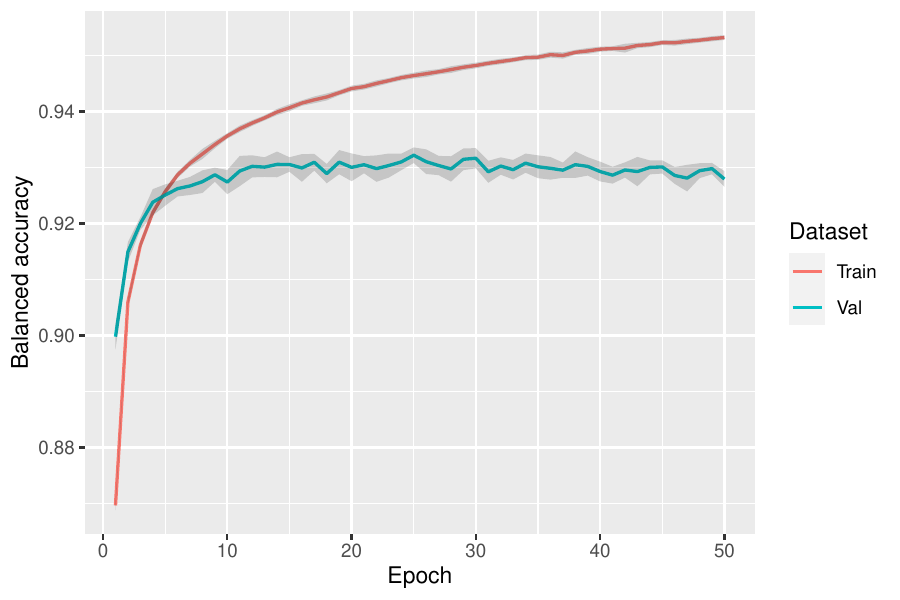}
        \caption{Balanced accuracy for tagger model.}
        \label{fig:tag_acc}
    \end{subfigure}
    \caption{By-epoch evaluation metrics for tagger network. Bands corresponding to a 95\% confidence interval are given in the shading surrounding the line plots. Note that the bands are present for the training metrics, but they are very tight.}
    \label{fig:tag_metrics}
\end{figure}

During training, the tagging network was evaluated with balanced accuracy calculated on the validation set. The sensitivity, specificity, and balanced accuracy are given in Table~\ref{tab:tag_acc}. The best balanced accuracy on the validation set was achieved after an average of 25.30 epochs ($SE = 2.65$). The by-epoch loss and balanced accuracy can be seen in Figure~\ref{fig:tag_metrics}.

\begin{table}[]
    \centering
    \begin{tabular}{lcccc}
        \toprule
         Data set & Loss & Sensitivity & Specificity & Balanced accuracy \\
        \midrule
         Train & $0.07 \pm 0.003$ & $0.95 \pm 0.002$ & $0.94 \pm 0.002$ & $0.95 \pm 0.002$ \\
         Validation & $0.08 \pm 0.002 $ & $0.92 \pm 0.005$ & $0.95 \pm 0.004$ & $0.94 \pm 0.001$ \\
         Test & $0.55 \pm 0.08$ & $0.85 \pm 0.02$ & $0.96 \pm 0.004$ & $0.90 \pm 0.009$ \\
         \bottomrule
    \end{tabular}
    \caption{Measures of accuracy for the model. Each value is given as its mean across the best epochs of the ten models trained $\pm 1.96\,SE$ to indicate a 95\% confidence interval estimate. Loss in this case was the weighted binary cross-entropy function, with a weight of 33 applied to the positive classes.}
    \label{tab:tag_acc}
\end{table}

The balanced accuracy in this model was substantially higher than the accuracy in the crisp model. This was to be expected since the data the model was trained on allowed ambiguous sounds to be classified more easily. Without actually assessing the boundaries, though, it is unclear whether this gain in accuracy will translate well to forced alignment boundaries.

\subsubsection{Alignment evaluation}

The neural networks qua acoustic models needed to be evaluated using more traditional (yet still task relevant) metrics. Yet, such evaluations do not directly correspond to how well a forced alignment system using the networks will perform. As such, the best network for each round was selected as the acoustic model for the alignment. We also trained the Montreal Forced Aligner on the TIMIT and Buckeye data set. The mean and median absolute errors of each network are given in Table~\ref{tab:abserr}.

\begin{table}[h]
    \centering
    \begin{tabular}{ccccc}
        \toprule
       Model & Interpolation & Data & \shortstack{Mean abs. \\ error (ms)} & \shortstack{Median abs. \\ error (ms)} \\
        \midrule
       Crisp & Yes & Train & $15.45 \pm 0.09$ & $6.59 \pm 0.05$ \\
       Crisp & Yes & Val & $14.82 \pm 0.15$ & $6.82 \pm 0.05$ \\
       Crisp & Yes & Test & $\mathbf{17.80 \pm 0.18}$ & $\mathbf{7.31 \pm 0.06}$ \\
        \midrule
        Crisp & No & Train & $16.55 \pm 0.07$ & $7.97 \pm 0.09$ \\
        Crisp & No & Val & $15.90 \pm 0.17$ & $8.21 \pm 0.11$ \\
        Crisp & No & Test & $18.91 \pm 0.2$ & $8.68 \pm 0.14$ \\
        \midrule
        Sparse & Yes & Train & $16.72 \pm 0.19$ & $7.29 \pm 0.09$ \\
        Sparse & Yes & Val & $15.35 \pm 0.15$ & $7.24 \pm 0.08$ \\
        Sparse & Yes & Test & $18.35 \pm 0.19$ & $7.79 \pm 0.1$ \\
        \midrule
        Sparse & No & Train & $17.99 \pm 0.21$ & $9 \pm 0.15$ \\
        Sparse & No & Val & $16.62 \pm 0.17$ & $9.04 \pm 0.11$ \\
        Sparse & No & Test & $19.67 \pm 0.22$ & $9.64 \pm 0.17$ \\
        \midrule
        MFA & --- & Train & 16.58 & 10 \\
        MFA & --- & Val & 16.47 & 10.01 \\
        MFA & --- & Test & 19.12 & 10.44 \\
       \bottomrule
         & 
    \end{tabular}
    \caption{Mean and median absolute error for the boundaries determined by the various forced alignment models. For the neural models, the error metrics are presented as the mean metric between all 10 rounds, plus or minus 1.96 time the standard error to provide a 95\% confidence interval. We only trained the Montreal Forced Aligner once, so there are no confidence intervals provided. The best performance on the hold-out test set in each column is indicated with boldface.}
    \label{tab:abserr}
\end{table}

We note that the Montreal Forced Aligner refused to train on and align a significant number of files and could not be made to do otherwise. In terms of boundaries, 46.76\% of the training boundaries were not produced, 1.72\% of the validation boundaries could not be produced, and 0.53\% of the test boundaries could not be produced. In terms of files, 52.37\% of training files, 0.10\% of validation files, and 0.06\% of test files were unable to be aligned. \footnote{We attempted to use a newer version of the Montreal Forced Aligner (v2.2.17) to see if these boundaries would align, but other fatal errors were encountered such that the training could not be started at all (corresponding to Issue 608 on the Montreal Forced Aligner GitHub repository).} As such, we can only interpret the data that was able to be aligned and offer a remark that there are sometimes mysterious, difficult to resolve issues when working with all forced alignment systems.

Overall, our best-performing network---the crisp model with interpolation---had less error than the Montreal Forced Aligner on all data sets. The Montreal Forced Aligner did perform better on some of the data sets than some of the neural nets we have trained, specifically the sparse network without interpolation. It is clear, though, that our neural networks categorically had better median absolute error. The best achieved error on the test data was from the neural networks overall as well. These results imply that the neural network acoustic models generally had better boundary placement, but the sparse networks tended to have more outliers than the Montreal Forced Aligner did.

It is interesting to note that the tagging approach to the label output did not consistently outperform the crisp, one-hot label approach. This result suggests that the tagging approach as implemented was not a great resolution to simulataneously meeting the \textsc{accurate classification criterion} and the \textsc{similarity reflection criterion}. It is possible that this is due to the empirical approach we employed to determine the tags, but it may also be that the tagging \textit{per se} is also an inappropriate resolution.

\begin{figure}
    \centering
    \includegraphics[width=0.75\textwidth]{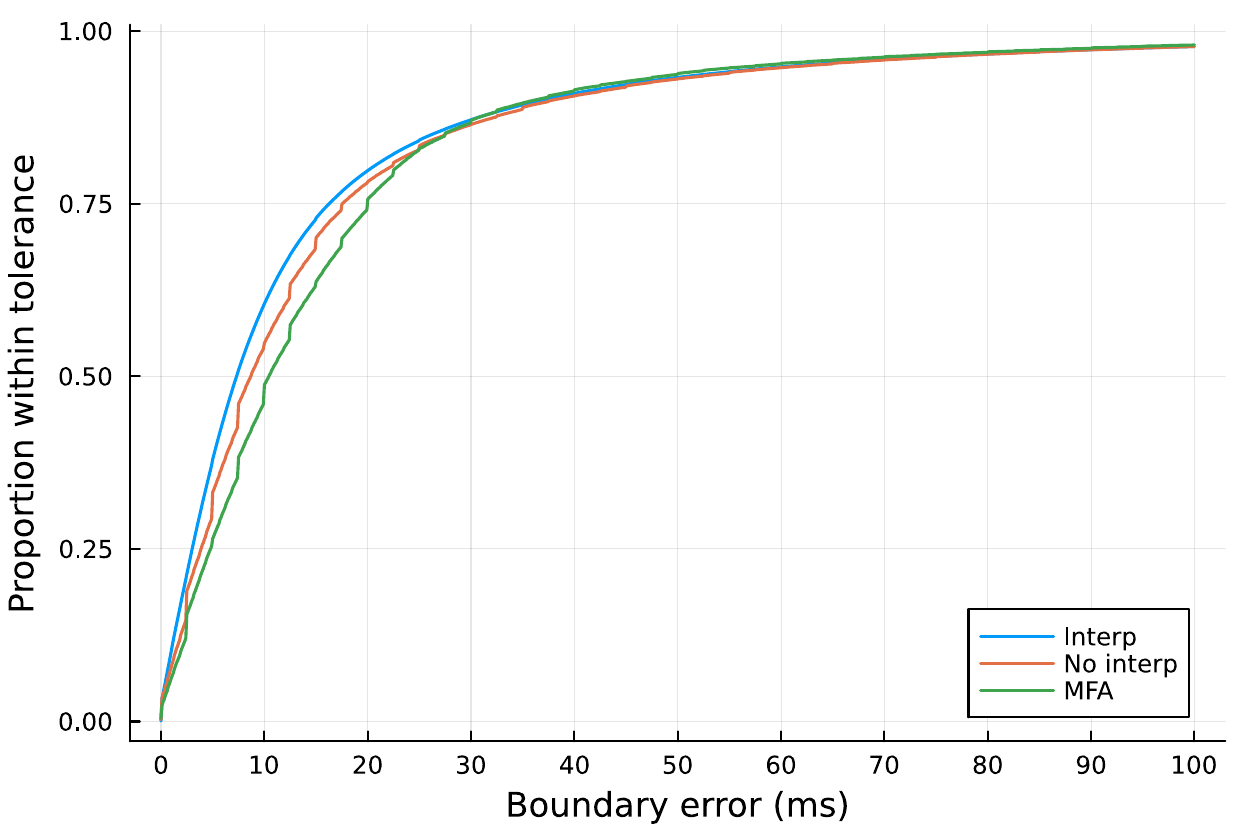}
    \caption{Cumulative density function for crisp networks and Montreal Forced Aligner. The line labeled ``Interp'' is the result of the system using interpolation. The line labeled ``No interp'' is the result of the system not using interpolation. The line labeled ``MFA'' is the result of the Montreal Forced Aligner.}
    \label{fig:cdf}
\end{figure}

The CDF for the crisp networks is presented in Figure~\ref{fig:cdf}. The plot indicates that the crisp networks tended to outperform the Montreal Forced Aligner in producing boundaries within approximately 25-30 ms of the target value, after which the Montreal Forced Aligner started to marginally outperform the crisp networks. Overall, it is arguable that boundaries with greater than 20 ms or 25 ms of error would need to be hand-corrected, so the crisp network, especially with interpolation, seems to provide some benefit by yielding the lowest errors within that tolerance.

We also created a plot for the sparse network results, which can be seen in Appendix~\ref{app:sparse_res}. It generally recapitulates the results in Table~\ref{tab:abserr} and Table~\ref{tab:proportions}. As such, there is not much additional descriptive information to obtain from it.

The proportion of values below specific error thresholds are presented in Table~\ref{tab:proportions}.\footnote{For the sake of readability, the confidence intervals are not presented here. A version of Table~\ref{tab:proportions} with confidence intervals is given in \ref{app:tol}.} At these specific values, the data suggest a similar conclusion as the boundary error statistics. Namely, the neural network acoustic models were on average more accurate but had more outliers than the Montreal Forced Aligner. That is, the neural network models had greater accuracy and wider variance, while the Montreal Forced Aligner had somewhat lower accuracy and tighter spread.

\begin{table}[]
    \centering
    \begin{tabular}{cccccccc}
        \toprule
         Model & Interpolation & Data & \multicolumn{5}{c}{Tolerance (ms)}  \\
         & & &  $< 10$ & $< 20$ & $< 25$ & $< 50$ & $< 100$ \\
         \midrule
         Crisp & Yes & Train & 64.12 & 82.78 & 86.84 & 94.87 & 98.48 \\
         Crisp & Yes & Val & 62.7 & 81.37 & 85.54 & 94.13 & 98.26 \\
         Crisp & Yes & Test & \textbf{60.48} & \textbf{79.8} & \textbf{84.19} & 93.31 & 97.79 \\
         \midrule
         Crisp & No & Train & 57.78 & 81.13 & 85.83 & 94.78 & 98.51 \\
         Crisp & No & Val & 56.61 & 79.74 & 84.6 & 93.99 & 98.26 \\
         Crisp & No & Test & 54.51 & 78.11 & 83.15 & 93.09 & 97.78 \\
         \midrule
         Sparse & Yes & Train & 60.45 & 80.14 & 84.7 & 94.09 & 98.29 \\
         Sparse & Yes & Val & 60.38 & 79.81 & 84.4 & 93.95 & 98.27 \\
         Sparse & Yes & Test & 58.11 & 78.17 & 82.9 & 92.99 & 97.81 \\
         \midrule
         Sparse & No & Train & 53.52 & 77.77 & 83.27 & 93.9 & 98.3 \\
         Sparse & No & Val & 53.44 & 77.42 & 83 & 93.68 & 98.25 \\
         Sparse & No & Test & 51.26 & 75.73 & 81.42 & 92.67 & 97.78 \\
         \midrule
         MFA & --- & Train & 49.63 & 77.9 & 86.03 & 95.88 & 98.86 \\
         MFA & --- & Val & 48.79 & 76.49 & 85.02 & 95.08 & 98.68 \\
         MFA & --- & Test & 47.28 & 74.74 & 82.91 & \textbf{93.81} & \textbf{98.02} \\
         \bottomrule
    \end{tabular}
    \caption{Error tolerances for the boundaries from the various alignment models. Each tolerance level is given in ms, and the result is given as a percent. The best test results for each column are indicated in bold.}
    \label{tab:proportions}
\end{table}

Overall, the region that is most likely of interest to phoneticians is within approximately 20 ms, and especially within 10 ms. In this region, the neural network aligner systems tended to outperform the Montreal Forced Aligner, whether interpolation was used or not. In the best case, our crisp model with interpolation had a 27.92\% relative increase of boundaries within the 10 ms threshold, and a 6.77\% relative increase of boundaries in the 20 ms threshold. Put more plainly, on the test set, the Montreal Forced Aligner had slightly less than 1 in 2 boundaries within the 10 ms threshold, while we have slightly more than 3 in 5 boundaries within the 10 ms threshold with our best system configuration. We also note that the percentage of boundaries within 20 ms is lower than in \citet{pengPracticalWayImprove2021}, but the training sets between the two projects are different, so the results are not entirely comparable.

\section{Discussion}

Of the methods we tested to improve the forced aligner system, only the interpolation solution we used seemed to have an appreciable effect on the boundaries the system generated. This is probably the most straightforward result we have. It is possible that future work could also make use of more sophisticated interpolation using polynomial or exponential bases. However, linear interpolation seems to provide decent results for little effort and compute.

The accuracy of the acoustic model is a particular conundrum. The tagging model had a much higher accuracy than the model used in the crisp model, but it did not translate to a very meaningful improvement in the boundaries the system calculated. A possible explanation for this finding is that the single-tier string of symbols imposed by the transcription style was not able to take advantage of the improved accuracy of the acoustic model. In that sense, the tagging acoustic model is somewhat at odds with the way in which the final transcription and boundaries must be formatted. In future improvements to our forced alignment system, we plan to explore other transcription representations that may permit the forced alignment system to benefit from the improved accuracy provided by the tagging approach. One such format may be multi-tiered segment timings, similar to gestural scores from articulatory phonology \citep{browmanArticulatoryPhonologyOverview1992}. This type of transcription has been alluded to in previous work as well \citep[][Chapter 6]{heselwoodPhoneticTranscriptionTheory2013}. Another potentially useful format could be one that provides a probability distribution for when a label is relevant or not, as could be determined from Bayesian neural networks or an ensemble of trained networks.

At a different level, there is the potential for concern about the acoustic features used. We did not try varying these from a standard speech recognition format. It is, theoretically, possible that a different set of features would have performed better in terms of segment classification. However, previous research that has configured neural network systems to learn acoustic features directly from the speech signal for segment recognition \citep{palazAnalysisCNNbasedSpeech2015,palazEstimatingPhonemeClass2013} and end-to-end speech recognition \citep{zeghidourEndtoEndSpeechRecognition2018} have not fared so well as to satisfy the \textsc{accurate classification criterion} significantly better than just using MFCCs. We believe such results are beginning to suggest that there may not be a set of acoustic features that would cause a typical acoustic model to satisfy the \textsc{accurate classification criterion}. That is, choosing to use MFCCs instead of some other set of acoustic features is not the limiting factor in satisfying the \textsc{accurate classification criterion}.

We also believe it is worth critically considering the \textsc{similarity reflection criterion}. The specification of the \textsc{similarity reflection criterion} was based on the imposition of segments on the speech signal, in addition to the structure of the segment types and articulatory parameters given in the IPA. It is certainly true that the IPA is not obligatorily a reflection of how humans perform speech communication. For example, it may seem from categorical perception results \citep{libermanDiscriminationSpeechSounds1957,libermanDiscriminationRelativeOnsettime1961,abramsonVoicetimingPerceptionSpanish1973} that humans do indeed have all-or-nothing responses and that all-or-nothing posterior distributions would be appropriate training targets. Beyond modern doubt about categorical perception \citep{mcmurrayMythCategoricalPerception2022}, we contend that this apparent reasonableness of all-or-nothing distributions is an illusion brought on by the necessity of ascribing a single label to a stimulus.

If the \textsc{similarity reflection criterion} were rejected in favor of all-or-nothing distributions, it would mean ignoring the acoustic similarities between different speech sounds, as we previously discussed. However, the moderate-to-high entropy in the ideal posterior distribution leads to a rather uncomfortable conclusion: Acoustics has limited informative value for segment identity, let alone word identity. This conclusion, of course, beggars belief, but doubt does little to resolve the apparent tension between satisfying the \textsc{similarity reflection criterion} and the informativity of speech acoustics. Indeed, some theories of linguistic communication hold that the purpose of a linguistic signal is to reduce ambiguity---and, thus, entropy---in the interlocutor \citep[e.g,][]{baayenComprehensionSegmentationProof2016}. If acoustics were truly almost uninformative, it would play almost no role in such theories, and acoustics would also have almost no useful role to play in theories that depend on recognizing segments from acoustics.

How, then, might this tension be resolved? We suspect that the resolution will require radical creativity. One potential solution is to change what is reflected about the speech signal by the categories that are presented for the acoustic model to learn. It may, for example, behoove the system to have symbols that rely on truly discriminative or contrastive moments in the speech signal. This type of symbology for the speech signal would make all-or-nothing posterior distributions more reasonable, though it might come at the steep cost of the transcription systems that phoneticians are familiar with. Investigating simultaneous segment identification, as we performed with our multi-label tagging model, may also prove to be a useful resolution. In fact, this type of system would allow for a relaxation of the tension between acoustic informativity and the \textsc{similarity reflection criterion}; acoustic informativity over the entire set of segments in a language would cease to be a concept altogether. The problem we previously outlined of how to rigorously determine what segments should be activated at what time remains to be resolved, however.

The overall task of forced alignment must also be considered critically. As we have outlined and hopefully demonstrated, there are a number of theoretical assumptions in forced alignment that make a high-performance system difficult to achieve. The corpora we have trained on have both had relatively detailed phonetic transcriptions and training an acoustic model that can account for these systems has been a perennial challenge. In the more general use case where a grapheme-to-phoneme conversion tool is used in forced alignment, the performance will almost assuredly be worse. The grapheme-to-phoneme tools will be based on some kind of static lexical representation, often of a citation form of a word. In the face of reduction and other processes that differentiate connected and conversational speech from citation forms, the boundaries a forced alignment system must come up with may be perniciously different from what a speaker has actually said.

We have also presented a model that was trained on very specific varieties of speech. TIMIT has some degree of dialectal variation represented among predominantly white speakers, and Buckeye has conversational speech balanced between white men and white women. Yet, it is starkly apparent that many common varieties of speech---such as Black English, L2 English, and English as spoken outside of the United States---are barely present or not present at all in the data we trained on. The propensity of machine learning models to overfit on the data they were trained on means it is all but certain that our model will perform worse on these varieties of speech not included in the training data, a common problem in automatic speech recognition \citep{wassinkUnevenSuccessAutomatic2022}. This is an inherent challenge in developing any kind of omnibus forced alignment system because it is nearly impossible to anticipate all possible research scenarios in which a tool will be used. Developing a training interface for forced alignment systems, as has been part of MFA since its inception \citep{mcauliffe_montreal_2017}, is an important method to mitigate some of these potential discrepancies, though some scenarios with small data make transfer learning difficult and from-scratch training all but impossible for neural networks. The supervised training method we used is also difficult to adapt to new data without already requiring segmented data.


Further work also needs to be performed on how best to evaluate forced aligners. Assessing boundary placement is a good starting place, and it is certainly a useful metric because the duration of segments and acoustic phenomena is a common measurement in phonetic research \citep{liskerCrosslanguageStudyVoicing1964,warnerIncompleteNeutralizationOther2004,podlubnyAssessingImportanceSeveral2018, perry_measuring_2023}. Researchers do frequently use acoustic measures beyond duration, however, and the boundaries can affect those measurements as well \citep{ahnOutlierAnalysisVowel2023, perry_modelling_2023, wangEvaulatingForcedAlignment2023}. Another relevant way the quality of a forced alignment system can be tested is through user preference. The boundaries a system generates can be presented to a trained human aligner, and the human ratings or adjustments to the boundaries can be quantified.

The code we used to train the system in the present paper is available in an archival GitHub repository at \url{https://github.com/MasonPhonLab/MAPS_Paper_Code/}. Additionally, an end-user system is being developed at \url{https://github.com/MasonPhonLab/MAPS}. The end-user system repository is where we will provide improvements and updates to our system. A model card \citep{mitchellModelCardsModel2019} for the model and system is also provided in the supplementary materials. The TextGrids that were created to evaluate the networks and the Montreal Forced Aligner are available in the George Mason University Dataverse at \url{https://doi.org/10.13021/orc2020/PDSAP7}.

\section{Conclusion}

Forced alignment systems still have a lot of room for improvement. We believe that using neural networks for the acoustic models is certainly promising, especially given their general ability to learn arbitrary function approximations. However, there are still many hurdles that must be cleared when creating and using such systems. Furthermore, the pattern recognition abilities of newer architectures like transformers \citep{vaswaniAttentionAllYou2017} as used and popularized in, for example, GPT-3 \citep{brownLanguageModelsAre2020} and GPT-4 \citep{openaiGPT4TechnicalReport2023} are both an exciting opportunity and a lurking problem. They are exceptionally good at learning sequential patterns. But, we would also like to put forth that pattern recognition is a double-edged sword in forced alignment; learning segmentation styles idiosyncratic to a particular data set is unlikely to be generally useful. Still, \citet{zhuPhonetoaudioAlignmentText2022} provided some hope that novel architectures may provide new avenues of development of aligners.

Nevertheless, we exhort the field to use forced alignment tools with a critical eye. The task itself may be ill-posed, representativeness in the training data is always suspect, and there are many areas where the component systems may not meet our expectations. Indeed, we believe it will be of benefit to the progression of forced alignment systems to imagine new tasks that might be more feasible for machines to perform. Perhaps even alternative representations of speech more amenable to current machine learning methods should be investigated as well.

\section{Acknowledgements}

This research was funded in part by SSHRC grant \#435-2014-0678 to the third author and the Kule Institute for Advanced Study through the Deep Learning for Sound Recognition group at the University of Alberta. We also thank the attendees of the 181st Meeting of The Acoustical Society of America for their feedback on an earlier version of this project. In addition, we would like to acknowledge the support of NVIDIA Corporation with the donation of a Titan X Pascal GPU and a Titan V GPU used for this research.

\section{References}

\bibliography{mybibfile}

\appendix

\section{Categorical cross-entropy gradient derivation}
\label{app:cce}

We begin by considering the form of softmax-activated logits through the categorical cross-entropy loss function. The categorical cross-entropy loss $\operatorname{cce}(z, y)$ calculated between the network's logit output vector $z$ and the target one-hot vector $y$ is

\begin{equation}
    \operatorname{cce}(z, y) = - \sum_i^n \psi_i \ln \left( \frac{e^{\zeta_i}}{\sum_j^n e^{\zeta_j}} \right) \,,
\end{equation}

\noindent where $n$ is the number of segment categories, $\psi_i$ is the $i$-th binary target from $y$, and $\zeta_i$ is the $i$-th logit from the network output vector before softmax activation $z$. Since $\psi_i$ is only nonzero for the target label (which only occurs once in the target vector $y$), we can simplify the expression to

\begin{equation}
    \operatorname{cce}(z, y) = - \ln \left( \frac{e^{\zeta_p}}{\sum_j^n e^{\zeta_j}} \right) \,,
\end{equation}

\noindent where $p$ is the index of the positive class in $y$.

Then, the partial derivative of $\operatorname{cce}(z, y)$ with respect to the logit output for the positive 
class $\zeta_p$ is

\begin{align}
    \frac{\partial \operatorname{cce}(z, y)}{\partial \zeta_p} &= \frac{\partial}{\partial \zeta_p} \left( - \ln \left( \frac{e^{\zeta_p}}{\sum_j^n e^{\zeta_j}} \right) \right) \\
    \label{eq:ccept1} &= - \frac{\partial}{\partial \zeta_p} \left(\zeta_p - \ln \left(\sum_j^n e^{\zeta_j} \right) \right) \\
    \label{eq:ccept2} &= - \left(1 - \frac{1}{\sum_j^n e^{\zeta_j}} e^{\zeta_p} \right) \\
    &= \frac{e^{\zeta_p}}{\sum_j^n e^{\zeta_j}} - 1 \,.
    \label{eq:apcce_corr}
\end{align}

The partial derivative of $\operatorname{cce}(z, y)$ with respect to the logit output for any particular negative class $\zeta_i$ where $i \neq p$, $\frac{\partial c}{\partial \zeta_i}$ is analogous to deriving $\frac{\partial c}{\partial \zeta_p}$. The result is

\begin{equation}
    \frac{\partial \operatorname{cce}(z, y)}{\partial \zeta_i} = \frac{e^{\zeta_i}}{\sum_j^n e^{\zeta_j}} \,.
    \label{eq:apcce_inc}
\end{equation}

\section{Binary cross-entropy loss gradient derivation}
\label{app:bce}

The binary cross-entropy loss $\operatorname{bce}(z, y)$ between the logit output vector of the network $z$ and the target vector $y$ is defined as

\begin{equation}
\operatorname{bce}(z, y) = - \sum_i^n{\left(\psi_i \ln \left (\frac{e^{\zeta_i}}{e^{\zeta_i} + 1} \right ) + (1 - \psi_i) \ln \left( 1 - \frac{e^{\zeta_i}}{e^{\zeta_i} + 1} \right)\right)} \,,
\end{equation}

\noindent and the partial derivative of $\operatorname{bce}(z, y)$ with respect to any particular logit output $\zeta_i$ is defined as

\begin{align}
\frac{\partial \operatorname{bce}(z, y)}{\partial \zeta_i} &= - \frac{\partial}{\partial \zeta_i} \sum_i^n{ \left( \psi_i \ln \left (\frac{e^{\zeta_i}}{e^{\zeta_i} + 1} \right ) + (1 - \psi_i) \ln \left( 1 - \frac{e^{\zeta_i}}{e^{\zeta_i} + 1} \right)\right)} \\
&= -\left( \psi_i \frac{\partial}{\partial \zeta_i} \ln \left (\frac{e^{\zeta_i}}{e^{\zeta_i} + 1} \right ) + (1 - \psi_i) \frac{\partial}{\partial \zeta_i} \ln \left( 1 - \frac{e^{\zeta_i}}{e^{\zeta_i} + 1} \right)\right) \\
&= - \left( \psi_i \frac{-1}{e^{\zeta_i}+1} + (1 - \psi_i) \frac{e^{\zeta_i}}{e^{\zeta_i}+1} \right) & \\
&= \begin{cases}
    \frac{-1}{e^{\zeta_i}+1} &\mbox{if } \psi_i = 1 \\
    \frac{e^{\zeta_i}}{e^{\zeta_i}+1} &\mbox{if } \psi_i = 0 \,,
    \end{cases}
\end{align}

\noindent where $\psi_i$ is the given target in $y$, a 0 or 1 value.

\section{Buckeye foldings}
\label{app:buckfold}

\begin{table}[H]
    \centering
    \begin{tabular}{cc}
    \toprule
         Buckeye symbol & Folded symbol  \\
         \midrule
         a & ah \\
    aan & aa \\
    aen & ae \\
    ahn & ah \\
    aon & ao \\
    awn & aw \\
    ayn & ay \\
    ehn & eh \\
    el & l \\
    em & m \\
    en & n \\
    eng & ng \\
    er & r \\
    ern & r \\
    eyn & ey \\
    h & hh \\
    hhn & hh \\
    ihn & ih \\
    iyn & iy \\
    nx & n \\
    own & ow \\
    oyn & oy \\
    tq & t \\
    uhn & uh \\
    uwn & uw \\
    $<$sil$>$ & sil \\
    \bottomrule
    \end{tabular}
    \caption{Foldings for Buckeye transcriptions. The left-hand column contains the original transcription symbols from Buckeye. The right-hand column contains the segment label they were folded into. The notation roughly corresponds to Arpabet.}
    \label{tab:buckfold}
\end{table}

\section{Tolerance results with confidence intervals}
\label{app:tol}

\begin{table}[H]
\centering
    \label{tab:tols_se_all}
    \hspace*{-0.75in}
    \begin{tabular}{llllllll}
        \toprule
         Model & Interpolation & Data & \multicolumn{5}{c}{Tolerance (ms)}  \\
         & & &  $< 10$ & $< 20$ & $< 25$ & $< 50$ & $< 100$ \\
         \midrule
         Crisp & Yes & Train & 64.12$\pm 0.24$ & 82.78$\pm 0.18$ & 86.84$\pm 0.16$ & 94.87$\pm 0.08$ & 98.48$\pm 0.02$ \\
         Crisp & Yes & Val & 62.7$\pm 0.27$ & 81.37$\pm 0.17$ & 85.54$\pm 0.17$ & 94.13$\pm 0.1$ & 98.26$\pm 0.05$ \\
         Crisp & Yes & Test & \textbf{60.48}$\mathbf{\pm 0.25}$ & \textbf{79.8}$\mathbf{\pm 0.18}$ & \textbf{84.19}$\mathbf{\pm 0.15}$ & 93.31$\pm 0.1$ & 97.79$\pm 0.05$ \\
         \midrule
         Crisp & No & Train & 57.78$\pm 0.36$ & 81.13$\pm 0.14$ & 85.83$\pm 0.13$ & 94.78$\pm 0.08$ & 98.51$\pm 0.02$ \\
         Crisp & No & Val & 56.61$\pm 0.38$ & 79.74$\pm 0.19$ & 84.6$\pm 0.15$ & 93.99$\pm 0.11$ & 98.26$\pm 0.05$ \\
         Crisp & No & Test & 54.51$\pm 0.44$ & 78.11$\pm 0.2$ & 83.15$\pm 0.16$ & 93.09$\pm 0.1$ & 97.78$\pm 0.05$ \\
         \midrule
         Sparse & Yes & Train & 60.45$\pm 0.42$ & 80.14$\pm 0.32$ & 84.7$\pm 0.28$ & 94.09$\pm 0.13$ & 98.29$\pm 0.05$ \\
         Sparse & Yes & Val & 60.38$\pm 0.35$ & 79.81$\pm 0.21$ & 84.4$\pm 0.19$ & 93.95$\pm 0.08$ & 98.27$\pm 0.04$ \\
         Sparse & Yes & Test & 58.11$\pm 0.42$ & 78.17$\pm 0.32$ & 82.9$\pm 0.28$ & 92.99$\pm 0.14$ & 97.81$\pm 0.06$ \\
         \midrule
         Sparse & No & Train & 53.52$\pm 0.55$ & 77.77$\pm 0.4$ & 83.27$\pm 0.33$ & 93.9$\pm 0.15$ & 98.3$\pm 0.05$ \\
         Sparse & No & Val & 53.44$\pm 0.48$ & 77.42$\pm 0.32$ & 83$\pm 0.24$ & 93.68$\pm 0.1$ & 98.25$\pm 0.04$ \\
         Sparse & No & Test & 51.26$\pm 0.57$ & 75.73$\pm 0.41$ & 81.42$\pm 0.32$ & 92.67$\pm 0.16$ & 97.78$\pm 0.06$ \\
         \midrule
         MFA & --- & Train & 49.63 & 77.9 & 86.03 & 95.88 & 98.86 \\
         MFA & --- & Val & 48.79 & 76.49 & 85.02 & 95.08 & 98.68 \\
         MFA & --- & Test & 47.28 & 74.74 & 82.91 & \textbf{93.81} & \textbf{98.02} \\
        \bottomrule
    \end{tabular}
    \caption{Error tolerances for the boundaries from the various alignment models, with the TIMIT and Buckeye data pooled together. Each number is presented as the mean over all 10 rounds, plus or minus 1.96 time the standard error to provide a 95\% confidence interval. The MFA results are an exception since the model was trained only once, so no confidence intervals are given.}
\end{table}

\section{Cumulative density function for sparse network}
\label{app:sparse_res}

\begin{figure}[H]
    \centering
    \includegraphics[width=0.75\textwidth]{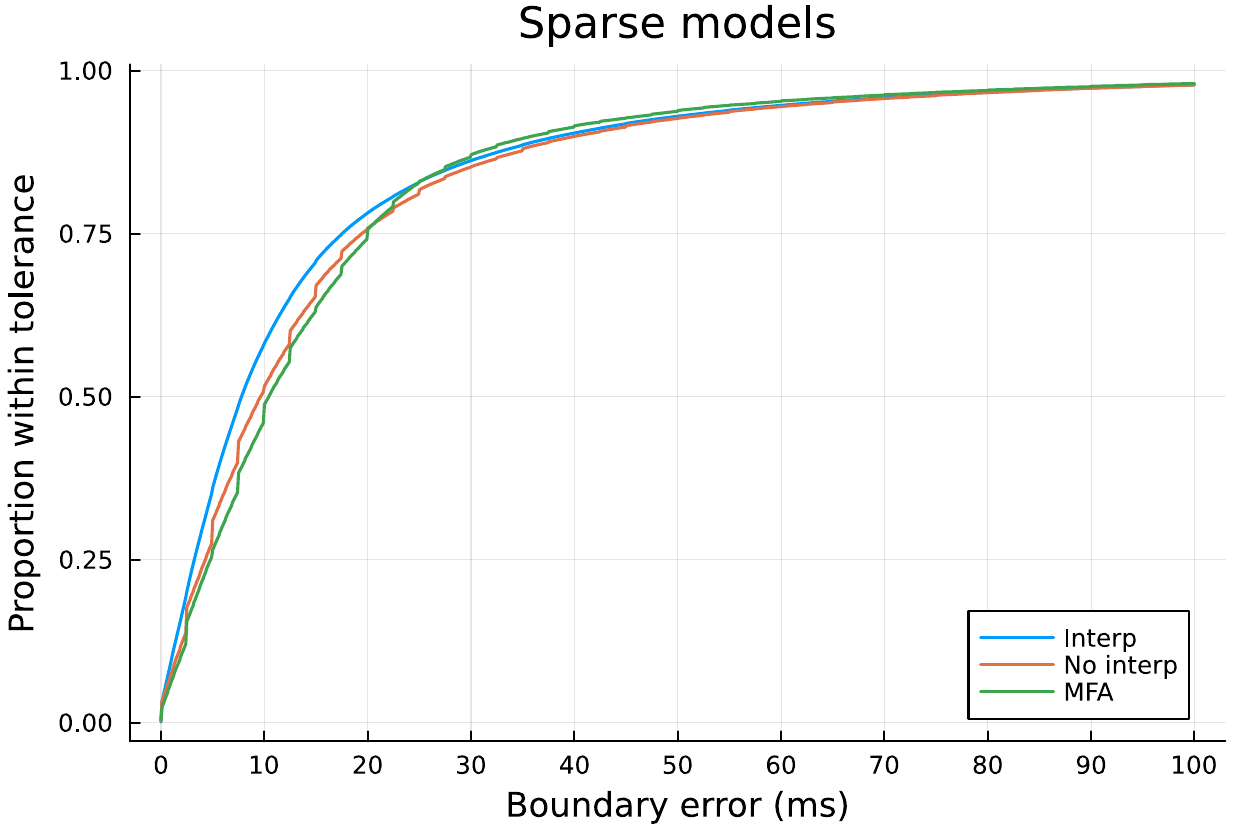}
    \caption{Cumulative density function for crisp networks and Montreal Forced Aligner. The line labeled ``Interp'' is the result of the system using interpolation. The line labeled ``No interp'' is the result of the system not using interpolation. The line labeled ``MFA'' is the result of the Montreal Forced Aligner.}
\end{figure}

\end{document}